\documentclass[twocolumn]{aa}
\usepackage{graphicx}
\usepackage{txfonts}
\usepackage{natbib}

\newcommand\ov{\over}
\newcommand\E[1]{\times10^{#1}}

\newcommand\U[1]{{\,\rm #1}}
\newcommand\kms{{km\,s^{-1}}}
\newcommand\cmq{{cm^{-2}}}

\newcommand\rs[1]{_{\mathrm{#1}}}       
\newcommand\RNFW{R\rs{NFW}}
\newcommand\rhoNFW{\rho\rs{NFW}}
\newcommand\zabs{z\rs{abs}}
\newcommand\zem{z\rs{em}}
\newcommand\mz{m_0}
\newcommand\Ncrit{N\rs{c}}
\newcommand\vgrav{V\rs{grav}}
\newcommand\vorb{V\rs{orb}}
\newcommand\vorbr{V\rs{orb,r}}

\newcommand\DvG{\Dl V_{\sigma}}
\newcommand\DvD{\Dl V_{\pm}}

\newcommand\HI{\hbox{H\,{\sc i}\,}}

\newcommand\MgI{\hbox{Mg\,{\sc i}\,}}
\newcommand\MgII{\hbox{Mg\,{\sc ii}\,}}
\newcommand\AlII{\hbox{Al\,{\sc ii}\,}}
\newcommand\AlIII{\hbox{Al\,{\sc iii}\,}}
\newcommand\SiII{\hbox{Si\,{\sc ii}\,}}
\newcommand\CrII{\hbox{Cr\,{\sc ii}\,}}
\newcommand\FeII{\hbox{Fe\,{\sc ii}\,}}
\newcommand\NiII{\hbox{Ni\,{\sc ii}\,}}
\newcommand\ZnII{\hbox{Zn\,{\sc ii}\,}}

\newcommand\NFeII{N(\hbox{\rm\hbox{Fe\,{\sc ii}\,}})}
\newcommand\NHI{N(\hbox{H\,{\sc i}\,})}

\newcommand\Dl{\Delta}
\newcommand\al{\alpha}
\newcommand\lmb{\lambda}
\newcommand\sg{\sigma}
\newcommand\om{\omega}
\newcommand\Om{\Omega}

\newcommand\Daova{\Dl\al/\al}
\newcommand\Dnuovnu{\Dl\nu/\nu}
\newcommand\Dnuovnui{\Dl\nu_i/\nu_i}
\newcommand\tmfv{\times10^{-5}}
\newcommand\avobs[1]{\langle#1\rangle\rs{obs}}

  \newcommand\iso[2]{\raise.5ex\hbox{\scriptsize #1}\rm #2}

\begin{document}
   \title{Can hidden correlations mimic a variable fine structure constant?}

   \titlerunning{Hidden correlations and variable fine structure constant}
   \authorrunning{R. Bandiera and E. Corbelli}

   \author{R. Bandiera
          \inst{1}
          \and
          E. Corbelli\inst{2}
          }

   \offprints{R. Bandiera,\\ \email{bandiera@arcetri.astro.it}}

   \institute{
              INAF-Osservatorio Astrofisico di Arcetri, Largo E. Fermi 5,
              I-50125 Firenze, Italy\\
              \email{bandiera@arcetri.astro.it}
         \and
              INAF-Osservatorio Astrofisico di Arcetri, Largo E. Fermi 5,
              I-50125 Firenze, Italy\\
              \email{edvige@arcetri.astro.it}
             }

   \date{Received 27 July 2004 / Accepted 20 December 2004}

   \abstract{
Murphy et al.\ (2003a, MNRAS, 345, 609) claim to find evidence of cosmological variations of the
fine structure constant $\alpha$ in the spectra of intervening QSO absorption
line systems.
We find that this result is affected by systematic effects.
The $\alpha$ values estimated in individual line systems depend on the set of
atomic transitions used and therefore the quoted dependence on the cosmic age
may reflect the fact that different sets of transitions are used at different
redshifts.
A correlation between line shifts and relative optical depths of the atomic
transitions is also present.
This correlation is very tight for a high-redshift subsample and accounts for
the anomalous dispersion of the $\alpha$ values found by Murphy et al.\ (2003a)
in this subsample.
The above correlations are consistent with a scenario in which gravitational
redshift, caused by the gravity of the dark halo, contributes to the shift in
frequency of individual components.
Gravitational redshift causes differential spectral shifts of the same order as
magnitude of those measured.
In the presence of line misidentification, these shifts can be interpreted in
terms of a variable $\alpha$.
In order to verify the gravitational redshift hypothesis, a direct access to
Murphy et al.\ (2003a) data, or to a large amount of new high resolution data,
is necessary.
   }

   \maketitle
%

\section{Introduction}

The hypothesis of a cosmological evolution of the fine structure constant $\al$
has been investigated in a series of papers \citep*[also collectively
referred to as MWF]{MWF03, WMF03,MWF01,WMF01,WF99}, by applying a sophisticated
spectral analysis, called the ``many multiplet'' method (hereafter MM), to large
samples of absorption line systems in QSO spectra.
This method is based on a $\chi^2$ analysis of multiple-component profile fits
in several transitions, and relies on theoretical calculations of how the
relative positions of different transitions change due to $\al$ variations
\citep[e.g.][]{DFK02}.
The use of many transitions improves the accuracy of $\al$ determination in
each absorption system, compared to fits of only two transitions \citep[like
the alkali-doublet method, e.g.][]{MWFad01}.
In addition, the use of large samples entails a much smaller quoted uncertainty
on the average value of $\Daova$ \citep[up to 128 absorption systems were
analyzed in][]{MWF03}.
The conclusion of MWF is that the value $\al$ was slightly smaller at earlier
cosmological epochs ($\Daova\sim-0.5\E{-5}$, with a 5-$\sg$ significance).
Ancillary papers \citep{MWFsys01,MWFsys03} have reached negative conclusions on
potential systematic effects which could mimic a variable fine structure
constant (hereafter VFSC).

A recent work by \citet{CSPA04}, based on more accurate measurements in high
quality data relative to a smaller sample objects, uses the MM method with
well-defined selection criteria.
No cosmological variation of $\Daova$ is found, with a 1-$\sg$ uncertainty of
$0.06\E{-5}$ (namely an order of magnitude below the values quoted by MWF).
\citet{CSPA04} avoid very weak and blended lines and point out that when one
does not do so, spurious detections are frequently seen.
Using a different but very robust method, \citet{BSS04} also do not find any
relevant time dependence of the fine structure constant.
\citet{BSS04} point out several inconsistencies in the results of MWF which
imply that systematic uncertainties due to misidentification of lines might be
significant in the MM method used by MWF.
In order to directly test the hypothesis of component misidentifications, one
needs to investigate in detail the original data and the routines used by MWF
in their analysis: this information is, however, not accessible to the
community.
As emphasized by \citet{CSPA04} and by \citet{BSS04}, MWF have not described
their algorithm of line identification, the confidence level adopted and the
physical assumptions made.
So, line misidentifications cannot be excluded in their results.

Given the overwhelming consequences in the fundamental physics and
cosmology of the variability of the fine structure constant \citep[see
e.g.][]{M03,U03,U04}, any newly suggested potential source of bias in the
determination of $\Daova$ must be examined in detail.
Direct access to MWF data and analysis would facilitate this process.
At the moment, however, we are forced to approach the problem in different
ways.
In the present paper we investigate the statistical properties of the data
presented by \citet[their Table~3]{MWF03}, and we find correlations that can
be explained only in terms of a systematic effect, not accounted for by MWF
(Section~2).
In Section~3, we put forward the hypothesis that differential gravitational
redshift (hereafter GRS), within absorption systems, may be responsible for the
correlations found and of the non-zero value of $\Daova$.
An important feature of GRS, with respect to other effects, is that symmetry
between red- and blueshifts is broken, and therefore a systematic effect on the
average values of the spectral shifts may be accounted for.
This effect is very tiny, and it can be detected only under particular
conditions.
In the MM analysis, GRS appears in only if misidentifications of some components
are present.
In Section~4, we search for correlations between $\alpha$ and some indicator of
the gravitational potential, for various absorption systems: the presence of
such correlations represents a strong clue that some differential redshift is
of gravitational origin.
Our last Section summarizes and concludes that GRS is a possible candidate for
the origin of the measured differential redshift.

\section{A statistical analysis of many-multiplet method results}

The central working hypothesis of MWF for evaluating $\Daova$ is the
proportionality between the line shift and the atomic relativistic correction
for all atomic transitions detected in the spectrum of each component of an
absorption system.
In the framework of the MM method, one expects the following relation to hold
for each atomic transition $i$:
\begin{equation}
\frac{\Delta\nu_i}{\nu_i}=Q_i\frac{\Delta\al}{\al}+K_z,
\end{equation}
where $\Dnuovnui$ are the line shifts measured for a specific component of the
absorption system, with their intrinsic sign.
The coefficients $Q_i$ are given by the ratio $q_i/\om_i$, where $\om_i$ is the
transition wavenumber, and $q_i$ is the atomic relativistic correction
\citep[see Table~2 of][]{MWF03}, and $K_z$ is the redshift of that specific
component.
If $\Dnuovnui$ and $Q_i$ are correlated, MWF analysis derives a value of
$\Daova$ different from zero.

We start with the working hypothesis that there is a primary correlation
between the relative optical depth of a transition and $\Dnuovnui$, i.e.\ that the
primary correlation is not that between $Q_i$ and $\Dnuovnui$.
A preliminary step is to estimate, for a ``typical'' absorber, the relative
optical depths of all transitions used by MWF.
We define the relative optical depth of the $i$-th transition, $\tau_i$, as the
optical depth scaled to that of a reference transition (from now on we
skip, for simplicity, the index $i$).
In this paper we will use \FeII\ at 2382\AA\ as the reference transition.
We take a sample of low-$z$ absorbers \citep{C97,CV01} to determine the column
density of \MgI\ and \MgII\ relative to that of \FeII, and a sample of high-$z$
absorbers \citep{PW99,PWT01} to determine the column density of \AlII, \AlIII,
\SiII, \CrII, \NiII, and \ZnII\ relative to that of \FeII.
We have linked the low and high-$z$ sets of data to derive the
relative optical depths of all transitions of interest.
Table~\ref{table1} gives the decimal logarithm of the relative optical depth,
$\log\tau$, the logarithm of the relative column density, $\log N\rs{rel}$,
the $Q$ coefficient, and some other basic data (ionic species, wavelength, and
oscillator strength $f$, as given by \citet{PWT01} for all transitions used.

\begin{table}
\begin{minipage}{84mm}
\caption{Atomic data.}
\begin{tabular}{@{}lrcrrr@{}}
\hline
\hline
\noalign{\smallskip}
Ion   &$\log N\rs{rel}$ &Wavelength &$f\times10^3$ &$\log\tau$
      &$Q\times10^2$                                           \\
\noalign{\smallskip}
\hline
\noalign{\smallskip}
\MgI   &$-1.6520$ &$2852.96$ \AA &$1810.0$ &$-0.8213$ & $0.2454$ \\
\MgII  &$ 0.3372$ &$2796.35$ \AA & $612.3$ & $0.6885$ & $0.5900$ \\
       &          &$2803.53$ \AA & $305.4$ & $0.3876$ & $0.3364$ \\
\AlII  &$-1.0454$ &$1670.79$ \AA &$1880.0$ &$-0.4306$ & $0.4511$ \\
\AlIII &$-1.5818$ &$1854.72$ \AA & $539.0$ &$-1.4641$ & $0.8606$ \\
       &          &$1862.79$ \AA & $268.0$ &$-1.7657$ & $0.4024$ \\
\SiII  &$ 0.4659$ &$1526.71$ \AA & $127.0$ &$-0.1288$ & $0.1038$ \\
       &          &$1808.01$ \AA &  $2.18$ &$-1.8207$ & $0.9601$ \\
\CrII  &$-1.6784$ &$2056.26$ \AA & $105.0$ &$-2.2263$ &$-2.2763$ \\
       &          &$2062.24$ \AA &  $78.0$ &$-2.3542$ &$-2.5799$ \\
       &          &$2066.16$ \AA &  $51.5$ &$-2.5336$ &$-2.7563$ \\
\FeII  &$ 0.0000$ &$1608.45$ \AA &  $58.0$ &$-0.9124$ &$-1.9301$ \\
       &          &$1611.20$ \AA &  $1.36$ &$-2.5415$ & $1.6918$ \\
       &          &$2344.21$ \AA & $114.0$ &$-0.4553$ & $2.9396$ \\
       &          &$2374.46$ \AA &  $31.3$ &$-1.0111$ & $3.8941$ \\
       &          &$2382.76$ \AA & $320.0$ & $0.0000$ & $3.5694$ \\
       &          &$2586.65$ \AA &  $69.1$ &$-0.6295$ & $3.9317$ \\
       &          &$2600.17$ \AA & $239.0$ &$-0.0892$ & $3.5259$ \\
\NiII  &$-1.4368$ &$1709.60$ \AA &  $32.4$ &$-2.5756$ &$-0.0342$ \\
       &          &$1741.55$ \AA &  $42.7$ &$-2.4477$ &$-2.4382$ \\
       &          &$1751.92$ \AA &  $27.7$ &$-2.6331$ &$-1.2263$ \\
\ZnII  &$-2.4081$ &$2026.14$ \AA & $489.0$ &$-2.2943$ & $5.0228$ \\
       &          &$2062.66$ \AA & $256.0$ &$-2.5676$ & $3.2528$ \\
\noalign{\smallskip}
\hline
\end{tabular}
\end{minipage}
\label{table1}
\end{table}

We define the standard deviation in $\log\tau$, and the linear correlation
between $\log\tau$ and $Q$ as:
\begin{eqnarray}
s    &=&\sqrt{\frac{\sum{(\log\tau-\avobs{\log\tau})^2}}{n\rs{obs}-1}},\\
\lmb &=&\frac{\sum{(Q-\avobs{Q})(\log\tau-\avobs{\log\tau})}}{\sqrt{
  \sum{(Q-\avobs{Q})^2}}\sqrt{\sum{(\log\tau-\avobs{\log\tau})^2}}},
\end{eqnarray}
where the sums are performed only over the $n\rs{obs}$ transitions in which a given
absorption system has been observed, and that have been used to derive $\Daova$
for that absorption system \citep[see Table~3 in][for the sets of transitions
used]{MWF03}.
Also the averaged values $\avobs{\log\tau}$ and $\avobs{Q}$ are computed using 
only the $n\rs{obs}$ transitions that have been observed.

The values of $s$ and $\lmb$ may change from one absorption system to another,
but there are absorption systems with equal values of $s$ and $\lmb$.
In principle, spectral lines should be weighted with their specific relevance
to the fit; however, in our simplified analysis we assume that all transitions
used for an absorption system are equally relevant.
The use of $\log\tau$, instead of $\tau$, has the advantage of limiting the
dispersion, even though the resulting correlations are qualitatively similar to
those obtained using $\tau$.
If $\log\tau$ correlates with $Q$, the primary correlation between $\log\tau$
and the measured $\Dnuovnu$ induces a spurious correlation between $Q$ and
$\Dnuovnu$.
In the MWF analysis this leads to a $\Daova$ different from zero.

Our complete sample is that listed in Table~3 of \citet{MWF03}.
We subdivide this into low-$z$ and high-$z$ subsamples, using a spectroscopic
criterion: an absorber is included in the high-$z$ sample only if it has been
observed in at least one transition at wavelengths shorter than 2300\AA.
\citet{MWF03} have noticed that a subsample of systems (called the
``high-contrast
sample''), observed in many transitions with very different relative optical
depths, has a statistical spread in $\Daova$ much larger than the average
nominal uncertainty of each absorber.
The authors ascribe this discrepancy to uncertainties not accounted for in
their analysis, but they exclude that this may lead to systematic effects in
the determination of $\Daova$.
We shall reach a different conclusion.

\begin{figure}
\centering
\includegraphics[width=75mm]{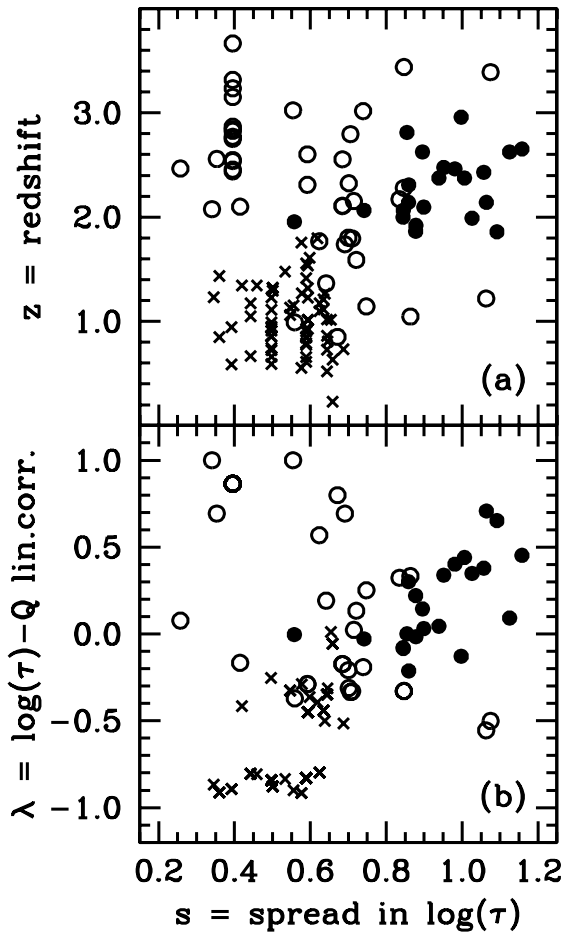}
\caption{
{\it(a)} Distribution of the data in the $s$-$z$ parameter plane and {\it(b)}
in the $s$-$\lmb$ plane.
Crosses are for objects in the low-$z$ sample and circles for the high-$z$
sample (filled circles indicate objects in the ``high-contrast'' subsample).
In {\it(b)} the number of points is smaller than in {\it(a)}, because in
{\it(b)} a single mark may indicate more than one absorption system.}
\label{fig:FIG1}
\end{figure}

Fig.~\ref{fig:FIG1} shows the complete set of data in the $s$-$z$ plane, and in
the $s$-$\lmb$ plane.
Fig.~\ref{fig:FIG1}{\it(a)} shows that our criterion, although purely based on
the set of transitions used, effectively separates lower and higher-$z$
absorbers, with a shallow cut around $z\sim1.8$.
Values of $s$ for low-$z$ absorbers are limited to the range ($0.3$,$0.7$);
values of $s$ for high-$z$ objects lie in the range ($0.3$,$1.2$).
gh-contrast absorption systems typically correspond to higher values of $s$
(say, above 0.8): this indicates that, at least on average, the quantity
$s$ accounts reasonably well for the spread in relative optical depth of the
atomic transitions used.

Because different sets of transitions are used for different redshifts, any
cosmological implication of measured variations of $\Daova$ with redshift
should be taken very cautiously.
Notice that, in addition, the low-$z$ and high-$z$ samples given by
\citet{MWF03} correspond to different kinds of absorbers.
The low-$z$ sample is mostly composed of absorption-line systems selected by
the \MgII\ lines,
and have moderate HI column densities; while a large fraction of the
high-$z$ sample is made up of damped Ly$\al$ absorption systems.
Therefore, we should not be surprised if, in our analysis, different results
are found for the two sets of systems, since they have been analyzed in
different sets of transitions, and furthermore are associated with different
types of cosmic structures.

In the two subsections below we will search for whether individual values of
$\Daova$ depend on $\lmb$.
A significant correlation would prove that our working hypothesis is 
correct; if instead the MWF results correspond to a true variation of $\al$ 
no correlation should be found.
For $\lmb$ positive, the (spurious) correlation between $\Dnuovnu$ and $Q$
will have the same sign as the primary correlation between $\Dnuovnu$ and
$\log\tau$.
For $\lmb$ negative the correlation will have the opposite sign.
In order to obtain a negative $\Daova$ when $\lmb$ is positive, $\Dnuovnu$
must decrease for increasing $\log\tau$.
When $\lmb$ is negative $\Dnuovnu$ must increase for increasing $\log\tau$.
According to our working hypothesis, the closer the $\lmb$ value is to $\pm1$, 
the larger the displacement from zero of the $\Daova$ value for each system.
Therefore we expect a correlation between $\Daova$ and $\lmb$.
The average value of $\Daova$ over the whole sample will be different from
zero, provided that the $\lmb$ distribution is not symmetric around zero.

\subsection{Low-$z$ sample}

Fig.~\ref{fig:FIG1}{\it(b)} shows that negative values of $\lmb$ are mostly
associated with low-$z$ objects (64 objects).
A weighted linear regression between $\Daova$ and $\lmb$ results in a
positive slope, although at a low significance level: $m=(0.80\pm0.60)\tmfv$,
with a reduced $\chi^2$ of 1.10 (assuming a functional dependence
$\Daova=m\lmb+q$).
A tighter result ($4.5\,\sg$) is obtained assuming a functional dependence
$\Daova=\mz\lmb$: $\mz=(0.92\pm0.20)\tmfv$, with a reduced $\chi^2$ of 1.09.
Below we again will use the symbols $m$ and $\mz$ to indicate the best fitting
slope leaving $q$ free, or setting $q=0$.

As can be seen in Fig.~\ref{fig:FIG1}{\it(b)}, low-$z$ objects present a
bimodal distribution, consisting of one component with $\lmb$ in the range
($-0.92$,$-0.80$), and of another one with $\lmb$ in the range
($-0.52$,$0.01$).
The former component is highly uniform in $\lmb$ (39 objects with an average
$\lmb$ value of $-0.85$) and the average $\Daova$ is $(-0.80\pm0.18)\tmfv$.
For the other component (25 objects with an average $\lmb$ value $-0.35$) we
have instead $\Daova=(-0.36\pm0.23)\tmfv$.
A comparison of the two subsamples again suggests that $\Daova$ increases with
$\lmb$.
However, as already obtained from the linear regression, the significance of
this trend becomes compelling only if we assume that $\Daova$ vanishes when
$\lmb=0$.

It is worth investigating another recently published work \citep{CSPA04}, which
analyses the spectra of a sample of low-$z$ absorption systems.
The authors do not find any evidence of a VFSC and give an upper limit of
$0.06\E{-5}$ to $\Daova$.
This is about one order of magnitude smaller than the signal claimed by MWF.
The sample of \citet{CSPA04} is smaller than that of \citet{MWF03} but the
quoted errors of $\Daova$ are smaller because high resolution observations
allowed them to select only systems with all individual components identified
unambiguously.
The decision of applying a severe selection of the systems could either render
the approach cleaner than that of MWF, or potentially more subject to biases.
Here we check whether the correlation between $\Daova$ and $\lmb$, found
in the MWF low-$z$ sample, is also present.

Values of $\lmb$ in \citet{CSPA04} were obtained using all transitions listed
for each of the 23 absorption systems in their Table~3 (even if some components
were analyzed by \citet{CSPA04} using a subsample of these transitions).
The estimated slope for the $\lmb$-$\Daova$ linear regression is
$m=(0.15\pm0.14)\E{-5}$, with a reduced $\chi^2$ of 0.94; in the case of a pure
linear proportionality we have $m_0=(0.12\pm0.08)\E{-5}$, with a reduced
$\chi^2$ of 0.91.
This result is consistent with no correlation between $\lmb$ and $\Daova$, down
to a 1.1--1.5~$\sg$ level, namely at a level 5--8 times smaller than the
correlation found in the \citet{MWF03} results.
We have also used different methods of object selection, such as limiting our
choice to systems in which both Fe and Mg transitions are detected in all
components (12 objects) or to systems in which the same set of transitions has
been used for all components (8 objects).
The conclusion, to a rather high level of confidence, is that the correlation
is not visible in the \citet{CSPA04} results.
Their method does not seem to be affected by the unwanted systematic effects
found in \citet{MWF03}.

\subsection{High-$z$ sample}

\begin{figure}
\centering
\includegraphics[width=84mm]{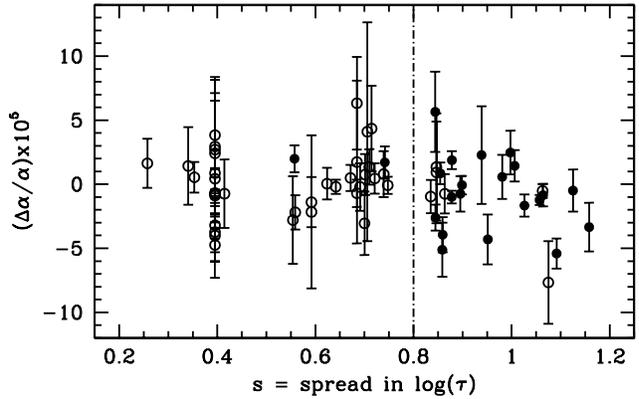}
\caption{
Dependence on the spread in the relative optical depths $s$ of the $\Daova$
values, for the high-$z$ sample.
The dot-dashed line divides the figure in two parts: for $s<0.8$ no trend of
$\Daova$ with $s$ is apparent; for $s>0.8$, instead, values of $\Daova$
decrease for increasing $s$.
As in Fig.~\ref{fig:FIG1}, filled circles indicate the ``high-contrast''
subsample.}
\label{fig:FIG2}
\end{figure}

Fig.~\ref{fig:FIG1}{\it(b)} shows that high-$z$ absorbers are distributed more
continuously over a wider range of $\lmb$, namely ($-0.6$,$+1.0$).
With respect to the low-$z$ sample, the high-$z$ sample shows a much larger
dispersion in $s$ (see Fig.~\ref{fig:FIG1}), and therefore it may be better
suited for searching for effects which arise when using transitions with a wide
range of relative optical depths (i.e.\ cases with large $s$ values).
Fig.~\ref{fig:FIG2} shows how the measured values of $\Daova$ depend on $s$ in
this sample.
While no apparent trend is present for $s<0.8$, for larger $s$ values there is
a clear trend of decreasing $\Daova$ with $s$.
A difference between these two subsamples is present also in the measured
values of $\Daova$: the average of $\Daova$ for $s<0.8$ (38 objects) is
$(-0.04\pm0.24)\tmfv$, for $s>0.8$ (26 objects) is $(-0.82\pm0.19)\tmfv$.
The discrepancy between these two values is at a $2.6\,\sg$ level.
A result similar to the $s>0.8$ sample, namely $\Daova=(-0.68\pm0.20)\tmfv$, is
obtained for the high-contrast sample: this is a rather obvious result, since
the two samples are partially overlapping.
In the remaining part of this section, we concentrate on a statistical
analysis of the high-contrast subsample.

The high-contrast sample is made of 22 objects for which \citet{MWF03} have
measured an excess dispersion of $\Daova$.
We propose that this excess dispersion results from the presence of hidden
variables, one of which is $\lmb$.
A simple way to derive how $\Daova$ varies with $\lmb$ is to evaluate the
average $\Daova$ for systems with negative and positive $\lmb$: we obtain
$\Daova$ equal to $(+0.63\pm0.43)\tmfv$ and $(-1.04\pm0.22)\tmfv$ respectively,
two values which are $3.5\,\sg$ apart.
A weighted linear regression between $\Daova$ and $\lmb$ on the high-contrast
sample gives a slope $m=(-3.46\pm0.88)\tmfv$ ($3.9\,\sg$ level).
Assuming that $\Daova$ vanishes at $\lmb=0$, the slope is
$\mz=(-3.21\pm0.62)\tmfv$ ($5.2\,\sg$ level).

In spite of the good significance level of $\Daova$-$\lmb$ regressions for the
high-contrast sample, their reduced $\chi^2$ values are still large, of the
order of 3.
A possibility is that the \citet{MWF03} hypothesis applies mostly to heavily
damped systems, because they are complex systems, containing many saturated
lines, while our conjecture of the presence of hidden variables may better
apply to absorption systems below a critical column density, $\Ncrit$.

\begin{table}
\begin{minipage}{84mm}
\caption{High-contrast absorption line systems.}
\begin{tabular}{@{}ccrcl@{}}
\hline
\hline
\noalign{\smallskip}
QSO & $\zabs$ & $\Daova$(10$^{-5}$)  & $\log\NHI$ & Ref.\footnote{
(L91) \citet{LWT91},
(L96) \citet*{LSB96},
(L98) \citet*{LSB98},
(O99) \citet*{OCC99},
(PW96) \citet{PW96},
(PW97) \citet{PW97},
(PW99) \citet{PW99},
(P01) \citet{PWT01},
(C02) \citet{CFK02}} \\
\noalign{\smallskip}
\hline
\noalign{\smallskip}
0100+13  & 2.310 & --3.941$\pm$1.368 & 21.40 & (PW99) \\
0149+33  & 2.140 & --5.112$\pm$2.118 & 20.50 & (PW99) \\
0201+37  & 1.955 &   1.989$\pm$1.048 & 20.20 & (PW96) \\
0201+37  & 2.462 &   0.572$\pm$1.719 & 20.40 & (PW96) \\
0528--25 & 2.141 & --0.853$\pm$0.880 & 20.70 & (L96) \\
0528--25 & 2.811 &   0.850$\pm$0.846 & 21.20 & (L96) \\
0841+12  & 2.374 &   2.277$\pm$3.816 & 20.95 & (PW99) \\
0841+12  & 2.374 &   1.435$\pm$1.227 & 20.95 & (PW99) \\
0841+12  & 2.476 & --4.304$\pm$1.944 & 20.78 & (PW99) \\
1011+43  & 2.959 &   2.475$\pm$1.706 & ..... & (C02) \\
1215+33  & 1.999 &   5.648$\pm$3.131 & 20.95 & (PW99) \\
1759+75  & 2.625 & --0.750$\pm$1.387 & 20.80 & (P01) \\
1759+75  & 2.910 & --0.492$\pm$1.645 & 19.80 & (O99) \\
1850+40  & 1.990 & --1.663$\pm$0.859 & ..... & (C02) \\
2206--20 & 1.920 &   1.878$\pm$0.702 & 20.65 & (PW97) \\
2230+02  & 1.859 & --5.407$\pm$1.179 &$<$20.85 & (PW99) \\
2230+02  & 1.864 & --0.998$\pm$0.492 & 20.85 & (PW99) \\
2231--00 & 2.065 & --2.604$\pm$1.015 & 20.56 & (PW99) \\
2231--00 & 2.065 &   1.707$\pm$1.249 & 20.56 & (PW99) \\
2231--00 & 2.653 & --3.348$\pm$1.904 &$<$20.30 & (L91)\\
2343+12  & 2.430 & --1.224$\pm$0.389 & 20.34 & (L98) \\
2359--02 & 2.095 & --0.068$\pm$0.722 & 20.70 & (PW99) \\
\noalign{\smallskip}
\hline
\end{tabular}
\end{minipage}
\label{table2}
\end{table}

\begin{figure}
\centering
\includegraphics[width=75mm]{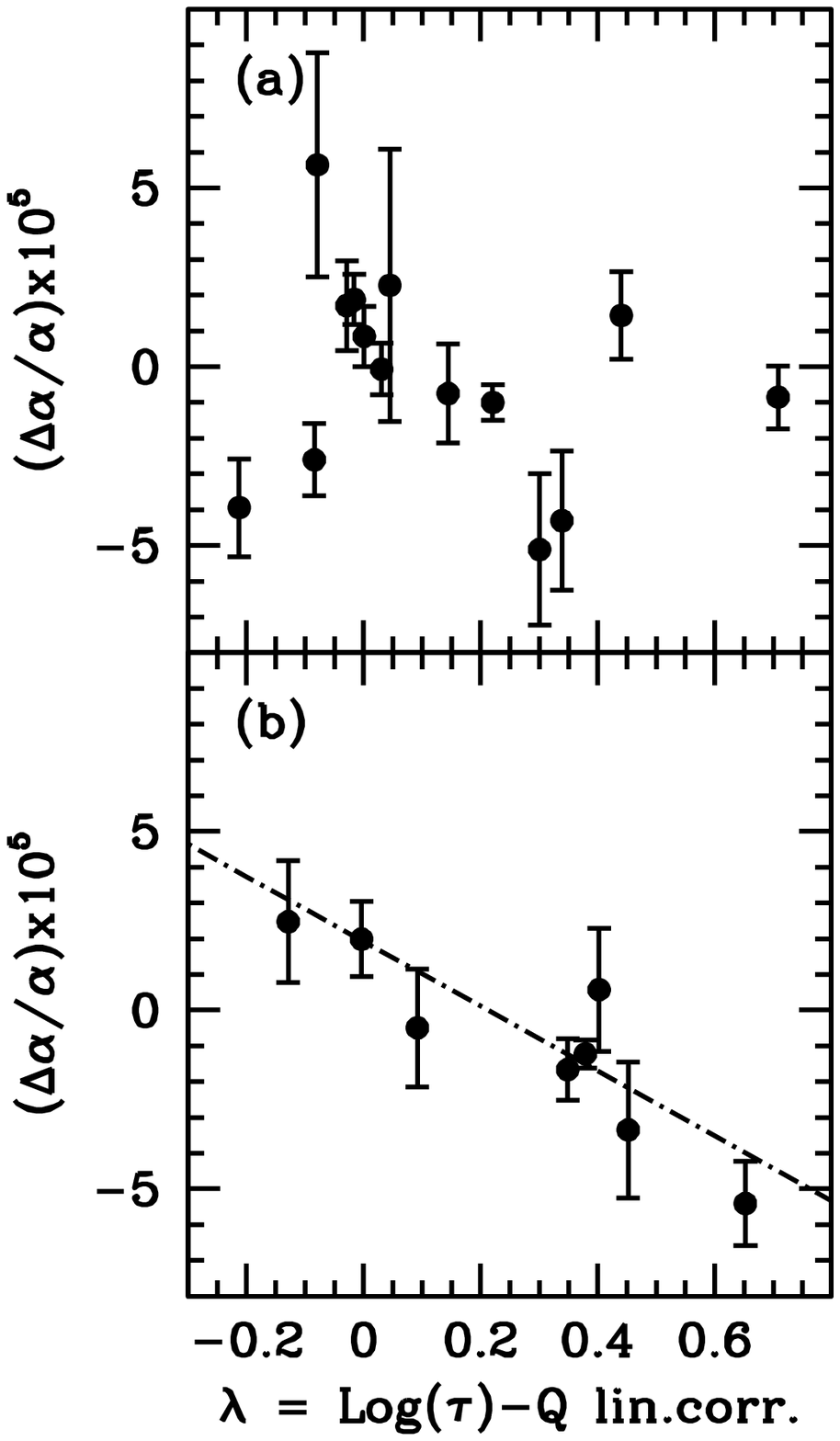}
\caption{
Correlations between $\Daova$ and $\lmb$ for high-contrast (high-$z$) objects,
{\it(a)} $\NHI>3\E{20}\U{\cmq}$, {\it(b)} $\NHI\le3\E{20}\U{\cmq}$.}
\label{fig:FIG3}
\end{figure}

In Table~\ref{table2} we list all high-contrast systems, and we indicate the
\HI\ column density whenever this can be found in the literature (the relative
references are listed in the last column).
For a few cases the \HI\ column density is unknown: we assume for those that the
\HI\ column density is below $\Ncrit$.
In Fig.~\ref{fig:FIG3}$(a)$ we show the data of Table 2 for
$\NHI>\Ncrit=3\E{20}\U{\cmq}$.
A weighted linear regression for these systems gives a slope
$m=(-1.08\pm1.15)\tmfv$ (with a reduced $\chi^2$ of 3.62); or alternatively
$\mz=(-1.42\pm0.97)\tmfv$ (with a reduced $\chi^2$ of 3.37).
Therefore there is no evidence of correlation in this subsample.
For $\NHI\le\Ncrit=3\E{20}\U{\cmq}$ instead a weighted linear regression gives
a slope $m=(-9.08\pm1.85)\tmfv$ (with a reduced $\chi^2$ of 0.91);
Fig.~\ref{fig:FIG3}$(b)$ displays this result.
Alternatively, $\mz=(-4.46\pm0.81)\tmfv$ (with a reduced $\chi^2$ of 1.87).
Here both regressions determine the slope at a significance level of about
$5\,\sg$.
The result we find is striking: the correlation between $\Daova$ and $\lmb$ is
very tight for low column density systems and it loses its significance for
heavily damped systems.
One worry about choosing $\Ncrit$ as low as $3\E{20}\U{\cmq}$ might be the 3
unknown \HI\ column densities of Table 2.
These can turn out to be higher than $3\E{20}\U{\cmq}$.
However, we have checked that as soon as $\Ncrit\le7\E{20}\U{\cmq}$ the
significance of the correlation stays above $4.5\,\sg$ for $\NHI\le\Ncrit$
(even though the dispersion increases) and stays below $1\,\sg$ for
$\NHI>\Ncrit$.
For $\Ncrit=7\E{20}\U{\cmq}$ for example, our last two statements are true not
only if all the unknown \HI\ column densities are below $\Ncrit$ but also if we
do not consider for the correlations the two entries relative to the (C02)
reference of Table 2, or if we consider one or both of them to be higher than
$\Ncrit$.

\section{Differential gravitational redshift in absorption systems}

In this section we discuss GRS in absorption systems, and its possible
influence on the quoted variation of the variable fine structure constant.
GRS can be detected in many types of bound systems such as stars, galaxies and
clusters of galaxies \citep[e.g.][]{BS00, SS93}, and has been used to infer
physical parameters like the total mass.
It is likely that absorption systems are localized in dark matter halos
of galaxies,
but since they can be observed only along one line of sight, whose position
with respect to the system barycentre is unknown, it is not easy to quantify
the magnitude of GRS in observable quantities.
To check explicitly if and how GRS plays a role in the MWF analysis is even
more difficult because their data and procedure are not available to the
public.

In order for GRS to be responsible for the fine structure constant variations
measured in absorption line systems with the MM method, three conditions are
required:\\
{\it i)} the differential GRS between the innermost regions and the outskirts
of typical absorption systems should be at least of the same order as the
relative shifts measured by MWF (of the order of $0.1\U{\kms}$, and interpreted
in terms of a VFSC);\\
{\it ii)} misidentification of individual components in a line profile
should affect the MM analysis, as carried on by MWF, i.e.\ components
attributed to the same cloud in different transitions are instead associated
with different clouds;\\
{\it iii)} the kinematics should allow a detection of a tiny spectral shift of
gravitational origin.

In the rest of this Section we examine these conditions in more detail.
In Section~4, we test in a more quantitative way the GRS hypothesis, by
searching for correlations between the estimated $\Daova$ and indicators of the
gravitational potential in the absorption system.

\subsection{Galactic haloes and gravitational redshift}

The aim of this subsection is to evaluate, for typical gaseous systems embedded
in dark matter haloes, the magnitude of the differential GRS.
Photons of frequency $\nu_0$ emitted in a gravitational potential $\Phi$ are
observed at infinity to be redshifted by an amount
\begin{equation}
{\Dl\nu\ov\nu_0}={\Phi\ov c^2},
\end{equation}
where $\Dl\nu=\nu-\nu_0$, and $c$ is the speed of light.

Let us consider gaseous systems embedded in dark matter haloes whose density
profile is described by the formula proposed by \citet*{NFW96,NFW97}:
\begin{equation}
\rho(R)=\frac{\rhoNFW}{\left({R/\RNFW}\right)\left(1+{R/\RNFW}\right)^2}
\label{eq:NFWprofile}
\end{equation}

\begin{figure}
\centering
\includegraphics[width=84mm]{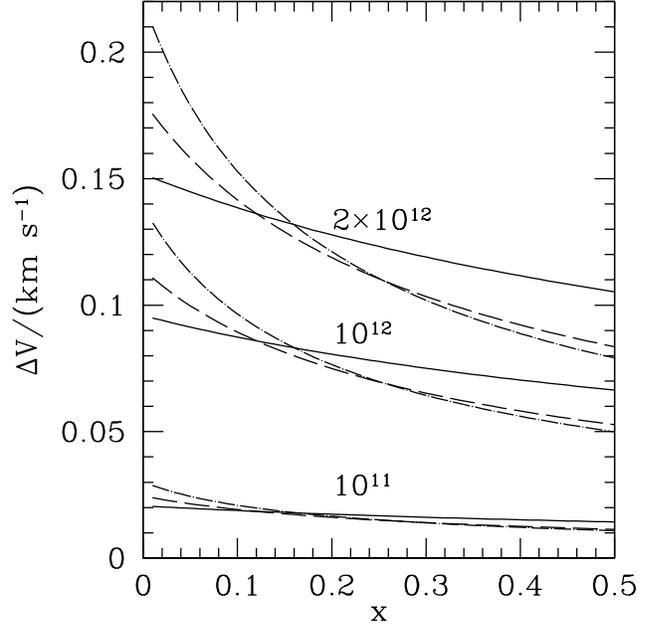}
\caption{
GRS in $\rm\kms$ as a function of the galactocentric distance $x=R/R_{200}$ at
$z=2$.
The 3 sets of curves correspond to the halo masses
$M_{200}h=10^{11},10^{12},2\E{12}\ M_{\odot}$.
The continuous lines are for concentration values $C=2$, the dashed lines for
$C=6$, the dash-dotted lines for $C=10$.}
\label{fig:FIG4}
\end{figure}

If $R_\Dl$ is the radius of a sphere containing a mean density $\Dl$ times the
critical density at redshift $z$, and $M_\Dl$ is the mass inside $R_\Dl$, one
can express the gravitational potential of a dark matter halo at a redshift $z$
in terms of $M_\Dl$ and of the concentration parameter $C_\Dl\equiv
R_\Dl/\RNFW$ \citep{NFW96,NFW97}.
Numerical simulations of structure formation often reveal a correlation between
$C_\Dl$ and $M_\Dl$ which depends on the details of the assumed cosmological
model \citep*[e.g. ][]{NFW96,NFW97,ACV01,ENS01}.
For a Cold Dark Matter-dominated universe, the dependence of $C_\Dl$ on
$M_\Dl$ is weak.
We consider $\Dl=200$, as originally proposed by \citet{NFW96,NFW97}, and
define $C\equiv C_{200}$.
Here we do not assume any explicit cosmological model to relate $C$ and
$M_{200}$ but use $C\simeq$2--10 at redshifts 0.5--3, as suggested by numerical
simulations of dark matter haloes forming in a Cold Dark Matter-dominated
universe.
The gravitationally redshifted frequency of an absorption line inside a
galactic potential dominated by dark matter can therefore be written as:
\begin{eqnarray}
{\Dl\nu\ov\nu_0}&=&-{G M_{200}\ov4\pi c^2R_{200}}
{\ln(1+Cx)\ov x\lbrace\ln(1+C)-C/(1+C)\rbrace}\nonumber \\
&\simeq&\!-{2.3\E{-8}\ov(\Om_0\Om)^{1/3}}\left({M_{200}/h^{-1}\ov
  10^{12}M_\odot}\right)^{2/3}\!\!\!\!\!\!
  {(1+z)\ln(1+Cx)\ov x\lbrace\ln(1+C)-C/(1+C)\rbrace},
\end{eqnarray}
where $G$ is the gravitational constant, $\Om$ is the matter density of the
Universe at $z$ and we have introduced the variable $x\equiv R/R_{200}$.
In Fig.~\ref{fig:FIG4} we show the GRS in $\rm\kms$ as a function of the
galactocentric distance $x$ at $z=2$ for $\Om_0/\Om\simeq 1$.
For a flat, cosmological constant dominated universe with $\Om_0=0.3$,
$\Lambda_0=0.7$, the GRS is a factor $\sim1.5$ larger than what
Fig.~\ref{fig:FIG4} shows.
There are no noticeable differences in the magnitude of the GRS if we consider
other gravitational potentials with similar masses but no cusp in the centre.
For a Burkert potential \citep{B95}, for example, we find less than a factor 2
difference over the whole range of masses and radii shown in
Fig.~\ref{fig:FIG4}.

An interesting result of Fig.~\ref{fig:FIG4} is that the velocity shifts are of
the same order of magnitude as those found in the analysis of \citet[see their
Table~4]{MWF03} and used to claim a VFSC.
The estimate of differential GRS presented above refers to the whole absorption
system; differential GRS within individual clouds (detected as individual
components in the line profile) is much smaller.

\subsection{Detectability of GRS in QSO absorption systems}

We outline here the most appropriate conditions to detect GRS effects in QSO
absorption systems.
In principle, GRS could be detected using even a single transition, if the
available sample of absorption lines is large enough to justify a statistical
approach.
In fact, isotropy arguments show that, in the absence of GRS, kinematic
redshifts of individual spectral components should average to zero with a
symmetric distribution in the whole sample.
GRS, instead, induces skewness in the spectral distribution of the absorption
components, because absorption lines located more deeply in the potential well
are those more strongly redshifted.
In this kind of analysis, the main practical difficulty is to cope with the
smallness of this effect, compared to both instrumental and statistical
uncertainties.
Even in the absence of instrumental uncertainties, there is a minimum number of
spectral components below which it is impossible to extract the gravitational
shift from the statistical dispersion due to kinematic motion.
The problem is qualitatively similar to that of determining, with accuracy
$\DvG$, the average shift for a sample of components whose distribution shows a
dispersion $\DvD$: in this case at least $\sim\left(\DvD/\DvG\right)^2$
different components are required.
If we aim at detecting $\DvG\sim0.1\U{\kms}$, while the dispersion of the
individual absorption components is $\DvD\sim100\U{\kms}$, at least $\sim10^6$
components are required, a number far beyond that available in the present data
sets.
This argument, even though in a more complex formulation, should apply also in
a multi-line approach.
A smaller number of components may be required if the velocity pattern of the
material changes smoothly with position.
For illustration, we present here a reasoning based on the
assumption that all blobs move in circular orbits, seen edge-on, in a
potential consistent with the profile given in Eq.~\ref{eq:NFWprofile}.
In this case, at a given radius $R$, the orbital velocity and the gravitational
redshift (expressed as a velocity) are given respectively by:
\begin{eqnarray}
\vorb^2&=&4\pi G\rhoNFW\RNFW^3 \nonumber \\
       &&\qquad\left(\ln(1+R/\RNFW)/R-1/(\RNFW+R)\right), \\
\vgrav&=&4\pi G\rhoNFW\RNFW^3\ln(1+R/\RNFW)/(Rc).
\end{eqnarray}
Let us also assume that, at $R=\RNFW$, the ratio $\vgrav/\vorb=1.577\sqrt{4\pi
G\rhoNFW}\RNFW/c$ is of the order of the observed ratio
$\DvG/\DvD\sim10^{-3}$.

\begin{figure}
\centering
\includegraphics[width=84mm]{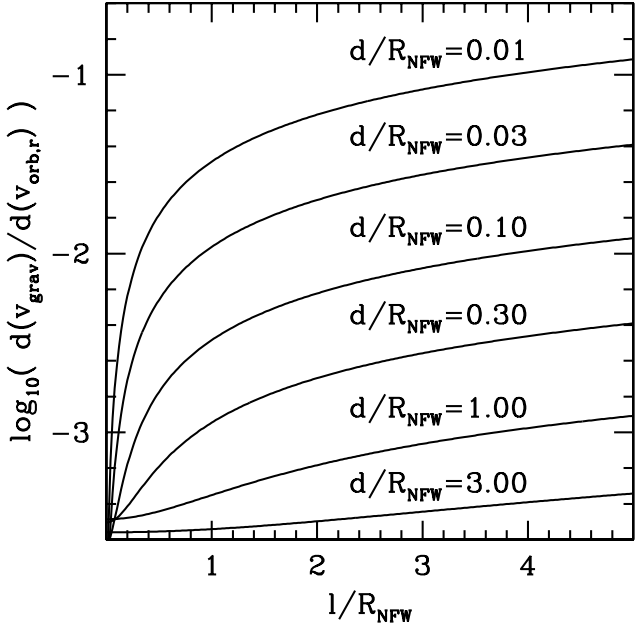}
\caption{
Relative change of $\vgrav$ and $\vorbr$, along the line of sight, for
different values of the impact parameter $d$.}
\label{fig:FIG5}
\end{figure}

We consider a path along a line of sight with impact parameter $d$ with respect
to the centre of the system: $R=\sqrt{l^2+d^2}$, where $l$ is the coordinate
along the line of sight.
We then compute $\vgrav$ and the radial component of $\vorb$ ($\vorbr$).
Results are displayed in Fig.~\ref{fig:FIG5} for different values of $d$.
While by a pure dimensional argument one would always expect values of the
order of $\DvG/\DvD\sim10^{-3}$, Fig.~\ref{fig:FIG5} shows that there may be
conditions in which this ratio can be one, or even two orders of magnitude
higher.
In addition, in the case of orbits tilted by an angle $\zeta$, a further
$1/\cos(\zeta)$ factor should be included.
This shows that, under favorable conditions, a sample of $\sim 100$ components,
spread over a set of absorption line systems, may be sufficient to get a
statistically significant detection of GRS.

\subsection{Component misidentification: how GRS breaks the symmetry}

MFW aim at comparing components, in different transitions, which are attributed
to the same cloud, i.e.\ to the same physical region within the absorption
system.
Of course, if this goal is reached, no GRS would be detected.
Let us examine instead the case of ``component misidentification'', namely the
case in which two components, identified in the line profile of two or more
different transitions, are erroneously attributed to the same cloud.
There may be different reasons why a given transition effectively samples inner
region clouds compared to another transition.
This may happen, for instance, if chemical or ionization radial gradients are
present in the absorption system.
There is some evidence of chemical and ionization inhomogeneities in high
column density absorbers, for example, which are very numerous in our high-z
sample \citep*{PS99,PSC02}.
But even in the absence of such gradients, inner regions are usually denser
than outer regions, and absorption lines of different relative optical depth
saturate at different distances from the centre and lines.
For this reason, lines with different $\tau$ might effectively sample different
regions in the potential well, and therefore may be subject to different levels
of gravitational redshift.
 From this scenario it naturally follows that one should find a correlation
between $\Dnuovnu$ and $\tau$ for different transitions, and therefore a
correlation between $\Daova$ and $\lmb$ (see Section~2).
Therefore our working hypothesis of Section~2 (a primary correlation between
$\Dnuovnu$ and $\tau$) is derived naturally from the idea that GRS and
component misidentification may play a role in some determination of
$\Daova$.
In our framework, if transitions of different relative optical depths sample
different regions of an absorption system, high-contrast absorption systems
should in fact be the most appropriate ones to reveal effects of differential
gravitational redshift.
The correlations we discuss in Section~2.2 indicate indeed that strong
systematic effects are present in this subsample.

In the presence of component misidentification, if clouds seen in absorption
in two different transitions are typically located at different distances from
the centre, their frequency ratio is affected by the different radial
velocities as well as by the different halo gravitational potential felt by the
two clouds.
Symmetry arguments imply that spurious deviations caused by the different
kinematic properties of the clouds average to zero in a suitably large sample.
Therefore only GRS can reasonably account for any residual systematic shift in
the average value of $\Daova$.

\section{Correlations between line frequency shifts and gravitational
potential}

In Section~2 we have shown that the concept of $\lmb$
is very effective in searching for hidden correlations.
This does not necessarily imply that GRS plays a role in generating
systematic effects, even though some working hypotheses suggest that it
might.
In this section we assume that, for a uniform sample in $\lmb$, the derived
value of $\Daova$ is roughly proportional to the level of GRS.
Since GRS is a function of the galactic potential, we investigate whether
$\Daova$ correlates with indicators of the gravitational potential strength.
The gravitational potential depends on the total mass associated with the
absorption system and on the radial distance of the absorbing material from the
centre of mass, but unfortunately none of these physical quantities is directly
measurable from absorption data.
A way out is to identify observable quantities, which correlate with total mass
and radius.

A quantity that can correlate with the strength of the gravitational potential
is the metal column density.
In fact, for constant metal abundances, more massive absorption systems have
higher gas column density (and therefore higher metal column densities).
In addition, metals are expected to be overabundant towards the centre of the
potential well or in more massive systems because of the longer star formation
history.

In this section we examine two samples of objects, both with known \FeII\ column
densities (hereafter $\NFeII$).
The first one contains low-$z$ absorption systems \citep[as listed in Table~3
of][]{MWF03} that have well documented $\NFeII$ along the line of sight.
Most of these objects have been analyzed by \citet[][C97]{C97}, or by
\citet[][CV01]{CV01}.
In order to obtain a uniform sample, we have limited our selection to these two
references.
The resulting sample contains 24 absorbing systems at redshifts $0.5<\zabs<
1.8$, and their $\log\NFeII$ is listed in Table~3.
This sample is statistically consistent with the larger low-$z$ sample analyzed
in the previous section.
Our second sample contains high-$z$, damped Ly$\al$ absorption systems (with
$\NHI\ge10^{20}$~cm$^{-2}$ and $\zabs<\zem$).
In order to have a nearly uniform sample we select 22 of them with
$1.8\le\zabs<3.5$, and published values of $\log\NFeII$ in at least one of the
following references: (L96) \citet{LSB96}, (PW96) \citet{PW96}, (PW97)
\citet{PW97}, (PW99) \citet{PW99}, (P01) \citet{PWT01}.
The sample is listed in the last 21 entries of Table~3.

\begin{table}
\begin{minipage}{84mm}
\caption{\FeII-selected absorption line systems.}
\begin{tabular}{@{}ccrcl@{}}
\hline
\hline
\noalign{\smallskip}
QSO & $\zabs$ & $\Daova$(10$^{-5}$) &$\log\NFeII$  &Ref.\footnote{
(C97) \citet{C97},
(CV01) \citet{CV01},
(L96) \citet{LSB96},
(PW96) \citet{PW96},
(PW97) \citet{PW97},
(PW99) \citet{PW99},
(P01) \citet{PWT01}} \\
         &       &                & cm$^{-2}$ &\\
\noalign{\smallskip}
\hline
\noalign{\smallskip}
0002+05  & 0.851 & --0.346$\pm$1.279 & 13.802 & (CV01)\\
0117+21  & 0.729 &   0.084$\pm$1.297 & 12.839 & (C97)\\
0117+21  & 1.048 & --0.223$\pm$2.200 & 12.718 & (CV01)\\
0117+21  & 1.343 & --1.290$\pm$0.948 & 12.484 & (C97)\\
0420--01 & 0.633 &   4.211$\pm$4.076 & 12.879 & (CV01)\\
0450--13 & 1.175 & --3.070$\pm$1.098 & 14.985 & (C97)\\
0454+03  & 0.860 &   0.405$\pm$1.325 & 15.068 & (CV01)\\
0454+03  & 1.153 & --0.749$\pm$1.782 & 12.826 & (CV01)\\
0823--22 & 0.911 & --0.394$\pm$0.609 & 13.597 & (CV01)\\
1148+38  & 0.553 & --1.861$\pm$1.716 & 13.186 & (CV01)\\
1206+45  & 0.928 & --0.218$\pm$1.389 & 12.905 & (CV01)\\
1213--00 & 1.320 & --0.738$\pm$0.760 & 14.529 & (C97)\\
1213--00 & 1.554 & --1.268$\pm$0.892 & 14.404 & (C97)\\
1222+22  & 0.668 &   0.067$\pm$1.474 & 13.115 & (CV01)\\
1225+31  & 1.795 & --1.296$\pm$1.049 & 13.881 & (C97)\\
1248+40  & 0.773 &   2.165$\pm$1.191 & 13.550 & (CV01)\\
1248+40  & 0.855 & --0.021$\pm$1.268 & 12.533 & (C97)\\
1254+04  & 0.519 & --3.371$\pm$3.247 & 13.957 & (CV01)\\
1254+04  & 0.934 &   1.485$\pm$1.908 & 12.699 & (CV01)\\
1317+27  & 0.660 &   0.590$\pm$1.515 & 13.100 & (CV01)\\
1421+33  & 0.843 &   0.099$\pm$0.847 & 13.143 & (C97)\\
1421+33  & 0.903 & --0.998$\pm$1.783 & 13.706 & (CV01)\\
1421+33  & 1.173 & --2.844$\pm$1.448 & 13.049 & (CV01)\\
1634+70  & 0.990 &   1.094$\pm$2.459 & 13.059 & (CV01)\\
\noalign{\smallskip}
\hline
\noalign{\smallskip}
0000--26 & 3.390 & --7.666$\pm$3.231 & 15.146 & (PW99)\\
0019--15 & 3.439 &   0.925$\pm$3.958 & 14.770 & (PW99)\\
0100+13  & 2.310 & --3.941$\pm$1.368 & 15.096 & (PW99)\\
0149+33  & 2.140 & --5.112$\pm$2.118 & 14.202 & (PW99)\\
0201+37  & 2.462 &   0.572$\pm$1.719 & 15.060 & (PW96)\\
0216+08  & 1.768 &   0.044$\pm$1.235 & 14.530 & (L96)\\
0347--38 & 3.025 & --2.795$\pm$3.429 & 14.503 & (P01)\\
0528--25 & 2.141 & --0.853$\pm$0.880 & 14.940 & (L96)\\
0741+47  & 3.017 &   0.794$\pm$1.796 & 14.041 & (P01)\\
0841+12  & 2.476 & --4.304$\pm$1.944 & 14.434 & (PW99)\\
1215+33  & 1.999 &   5.648$\pm$3.131 & 14.648 & (PW99)\\
1425+60  & 2.827 &   0.433$\pm$0.827 & 14.480 & (L96)\\
1759+75  & 2.625 & --0.750$\pm$1.387 & 15.091 & (P01)\\
1946+76  & 2.843 & --4.743$\pm$1.289 & 13.380 & (L96)\\
2206--20 & 1.920 &   1.878$\pm$0.702 & 15.458 & (PW97)\\
2206--20 & 2.076 &   1.429$\pm$3.022 & 13.320 & (PW97)\\
2230+02  & 1.864 & --0.998$\pm$0.492 & 15.188 & (PW99)\\
2231--00 & 2.066 & --2.604$\pm$1.015 & 14.750 & (PW99)\\
2348--14 & 2.279 &   1.346$\pm$4.180 & 13.792 & (PW99)\\
2359--02 & 2.095 & --0.068$\pm$0.722 & 14.507 & (PW99)\\
2359--02 & 2.154 &   4.346$\pm$3.338 & 13.895 & (PW99)\\
\noalign{\smallskip}
\hline
\end{tabular}
\end{minipage}
\label{table3}
\end{table}

In order to estimate the dependence of the measured values of $\Daova$ on
$\NFeII$, we will use a linear regression, adopting the functional dependence
$\Daova=m(\log\NFeII-13.5)+q$.
In this way, for the whole low-$z$ sample of Table~3, we obtain
$m=(-0.45\pm0.33)\tmfv$ and $q=(-0.50\pm0.25)\tmfv$, with a reduced $\chi^2$ of
0.90: the value obtained for the the slope $m$ is not statistically significant
($1.4\,\sg$).
For the high-$z$ sample of Table~\ref{table3}, instead, we calculate
$m=(1.19\pm0.51)\tmfv$ and $q=(-2.21\pm0.72)\tmfv$, with a reduced $\chi^2$ of
2.65: a reasonable ($2.3\,\sg$) significance level is obtained here, but the
reduced $\chi^2$ is anomalously high.
We attribute this result to the fact that some of the quoted uncertainties on
$\Daova$, which are relative to high-contrast damped systems, are
underestimated (see Fig.~\ref{fig:FIG3}{\it(a)}, and the discussion in
the previous section).
In fact, if we compute a linear regression on the high-$z$ sample, assigning
the same uncertainty to all measurements of $\Daova$, we obtain
$m=(-0.38\pm1.20)\tmfv$ and $q=(-0.27\pm1.45)\tmfv$: therefore, also in this
case the slope $m$ is not statistically significant.
We attribute the low significance of these correlations to the fact that none
of the two samples is uniform in $\lmb$.

Luckily, 18 out of the 24 objects of the low-$z$ sample lie in a narrow range
of $\lmb$, namely $(-0.92,-0.79)$.
The correlation becomes more significant if we select only absorbers with high
$\log\NFeII$ values in this $\lmb$ range.
Taking $\log\NFeII>13.0$ (12 objects) we obtain $m=(-1.23\pm0.56)\tmfv$
($2.2\,\sg$) and $q=(-0.35\pm0.34)\tmfv$, with a reduced $\chi^2$ of 1.14.
For $\log\NFeII>13.5$ (6 objects) $m=(-2.16\pm0.78)\tmfv$ ($2.8\,\sg$) and
$q=(0.34\pm0.53)\tmfv$, with a reduced $\chi^2$ of 0.92.
This result is consistent with a $\Daova$ that deviates significantly from zero
only for $\log\NFeII$ larger than about $13.5$.
The $\Daova$ deviation is towards more negative values as $\log\NFeII$
increases.
Although the statistical significance is not high (only 2.8$\sigma$ in the best
case), one should notice that negative deviations of $\Daova$ for negative
$\lmb$ are consistent with the increase of $\Daova$ with increasing $\lmb$,
obtained in the previous section for the low-$z$ sample.
This can be explained in terms of an effective line segregation, with weaker
lines sampling effectively regions closer to the centre of the potential well.
GRS is a reasonable candidate to explain the decrease in $\Daova$ with
increasing $\log\NFeII$, because systems with higher metal column density are
either closer to the centre of the potential well, or located in more massive
systems or both.

There have been other proposals to explain the VFSC result.
Of particular interest is the consistency of the VFSC values with a non-solar
isotopic ratio of $(\iso{25}{Mg}+\iso{26}{Mg})/\iso{24}{Mg}$ occurring at large
redshifts.
If this isotopic ratio increases at large redshifts, small apparent shifts
would be introduced in the absorption lines \citep*{AMO04}.
These shifts mimic a VFSC if instead a solar isotopic ratio of magnesium is
used, as in the MWF analysis.
The large uncertainty about this proposal lies in the magnesium isotopic ratio
variations with metallicity.
\citet{AMO04} show that an Initial Mass Function (IMF) particularly rich in
intermediate mass stars is needed for an increase of the magnesium isotopic
ratio with decreasing metallicity (in the range [Fe/H]=0, $-1.5$).
Analyses of low metallicity star data \citep{GL00} show instead an opposite
trend of the magnesium isotopic ratio variations with metallicity.
If the data by \citet{GL00} reflect the situation in QSO absorbers, which have
lower metallicity than solar, then magnesium isotopic ratio variations cannot
explain the negative $\Daova$ reported by MWF.
In this case metallicities lower than solar would correspond to lower magnesium
isotopic ratios and to positive value of $\Daova$.
If one assumes instead that the results of \citet{AMO04} apply to gas in QSO
absorbers, then magnesium isotopic ratio variations can explain the MWF
results.
The correlation between $\Daova$ and metallicity should in this case be
positive for a certain range of [Fe/H] (which depends on IMF and yields).
For the GRS hypothesis instead large variations of $\Daova$ should be
associated with large potential wells.
Since more massive galaxies have had a longer star formation history, a negative
correlation between metallicity and $\Daova$ should appear.
The results of this Section seem to favour the GRS hypothesis, but accurate
metal abundances, higher significance in the outlined correlations and more
information on earlier IMF are needed to draw any conclusions on the magnesium
isotopic ratio variation hypothesis.

\section{Conclusions}

Even though we cannot disprove completely the VFSC hypothesis of \citet{MWF03},
in
this paper we have discovered systematic effects which can mimic a non-zero
$\Daova$.
Often $\Daova$ depends on $\lmb$, the correlation coefficient between two
atomic quantities: the relative optical depth and the relativistic correction
coefficient $Q$.
For the atomic transitions used in the MWF analysis for each absorber, we
evaluate $\lmb$ and notice that the $\lmb$ distribution is not symmetric around
zero: this justifies the net displacement of the average $\Daova$ from zero.
In particular, the correlation between spectral shifts and the relative optical
depth of the various transitions seems to be the origin of correlations between
spectral shifts and $Q$, used by MWF as evidence for a VFSC.
Non-zero values of $\Daova$ may then be the result of GRS if lines of different
relative optical depth effectively sample regions at different distances from
the centre of the potential well.

We have examined the low-$z$ and the high-$z$ samples of \citet{MWF03}
separately, since they have been analyzed in different sets of transitions and
furthermore are associated with different types of cosmic structures (see
Section~2 for more details).
In the low-$z$ sample the correlation between $\Daova$ and $\lmb$ can be
interpreted in terms of a primary positive correlation between $\Dnuovnu$ and
the relative optical depth: that means that relatively weaker lines are more
redshifted than others (i.e.\ $\Delta\nu$ more negative).
This is in agreement with what one would expect from GRS, under the hypothesis
that intrinsically weaker lines are visible only through dense regions, located
closer to the centre of the absorption system.
Since for this sample most of the systems lie in a restricted $\lmb$ range, and
we know the metal column densities, we compare $\NFeII$ with the derived
$\Daova$ values.
Systems with higher $\NFeII$ values have more negative $\Daova$ values.
This again confirms the GRS hypothesis since more negative $\Daova$ can be
interpreted as more heavily gravitationally redshifted lines.
High $\NFeII$ values are in fact expected when the gas lies in regions of
strong gravitational potential.
If future data prove a positive correlation between $\Daova$ and metal
abundances, this will support GRS-related studies and weaken other
proposals, such as the magnesium isotopic ratio variations, for explaining the
non-zero value of $\Daova$ derived in the MWF analysis.

For the ``high-contrast'' subsample of the high-$z$ systems, for which
\citet{MWF03} have reported an anomalous statistical dispersion in $\Daova$, we
find instead a systematic effect that can be interpreted in terms of a primary
inverse correlation between $\Dnuovnu$ and the relative optical depth.
Further dividing this subsample into heavily damped systems, with
$\NHI>\Ncrit$, and systems with $\NHI\le\Ncrit$, we have discovered that the
latter subsample shows a striking inverse correlation, at a level higher than
$4.5\,\sg$ for $\Ncrit\le7\E{20}\U{\cmq}$.
No correlation is present for $\NHI>\Ncrit$.
So, while for the heavily damped systems the MWF interpretation of the
anomalous statistical dispersion might be correct, for lower column density
systems we believe that the anomalous statistical dispersion is indeed the
effect of an underlying correlation.

Note that the sign of the correlation between $\Daova$ and $\lmb$ for the
high-contrast (high-$z$) sample is opposite to that obtained for the low-$z$
one.
If this correlation is interpreted in terms of GRS, stronger transitions should
be associated with innermost regions of the absorption system.
This is opposite to what is inferred for the low-$z$ sample, and requires the
presence of physical inhomogeneities within the system \citep[as actually found
in some cases. e.g.][]{PS99,PSC02}.
However, one cannot exclude that the opposite trends simply arise because the
bias comes out differently in the MWF procedure used to estimate $\Daova$:
this is because the data and the characteristics of the absorption systems in
the two samples are very different.

As already pointed out, a more detailed analysis can be carried out having
access to the original data.
For instance, it would be useful to obtain subsamples analyzed uniformly, using
the same set of transitions.
Conversely, one could also investigate how $\Daova$ changes with different
choices of the fitted transitions, in the same absorption system.
More generally, instead of determining how $\Dl\nu/\nu$ correlates with $Q$
(from which $\Daova$ is determined), one should search for correlations of
$\Dl\nu/\nu$ with other quantities, such as the relative optical depth.
Notice that any quoted variation of $\Daova$ with cosmological time should be
taken very cautiously, because different sets of transitions are used for
different redshifts.

In a recent work, \citet{CSPA04} found no variation in the fine structure
constant for a sample of low-$z$ absorption systems.
They obtained this result avoiding both weak and saturated lines, as well
as heavily blended spectral regions.
In this way they restricted their analysis mostly to intermediate
``satellite components'', avoiding the ones relative to central or outermost
regions.
Therefore they reached a better accuracy in positioning the
subcomponents.
A contamination of GRS on $\Daova$ is possible only in the
presence of some bias in the spectral analysis.
We do not find any significant correlation between $\Daova$ and $\lambda$ in
the \citet{CSPA04} results.
Their work further supports the idea that biases may originate from the fit of
weak spectral components or of the saturated and complex parts of a line
structure.
Moreover, GRS effects are harder to detect in a sample which avoids the
centremost and outermost regions since this choice effectively restricts the
range of the potential well tested.

We would like to encourage any future GRS experiment over
cosmological distances since the results can be unique tools to test
structure formation scenarios in the context of several cosmological models.
For example if at redshifts as high as 4--5 the Universe is still dominated by
small mass objects (Cold Dark Matter scenario), GRS-induced line frequency
shifts in absorption systems at such redshifts should be smaller than the ones
detectable at more recent times.
Also theoretical models of possible variations of the fine structure constant,
which take into account the observational state of the art are needed
\citep[see
e.g.][]{S03}, in order to check the consistency of different results.

\begin{acknowledgements}
\end{acknowledgements}

We are grateful to the referee and to Charles Steinhardt for many valuable
comments to the original version of this manuscript.


\end{document}